\documentclass[sigconf,authorversion]{acmart}

\AtBeginDocument{%
  }

\copyrightyear{2025}
\acmYear{2025}
\setcopyright{rightsretained}
\acmConference[MuC '25]{Mensch und Computer 2025}{August 31-September 03, 2025}{Chemnitz, Germany}
\acmBooktitle{Mensch und Computer 2025 (MuC '25), August 31-September 03, 2025, Chemnitz, Germany}
\acmDOI{10.1145/3743049.3748581}
\acmISBN{979-8-4007-1582-2/25/08}


\usepackage{subcaption}

\begin{document}

\title{Pull Requests From The Classroom: Co-Developing Curriculum And Code}

\author{Dennis Zyska}
\orcid{0009-0003-4914-1015}
\affiliation{%
  \institution{Ubiquitous Knowledge Processing Lab (UKP)\\
  Technical University of Darmstadt}
  \city{Darmstadt}
  \country{Germany}
  }
\email{dennis.zyska@tu-darmstadt.de}

\author{Ilia Kuznetsov}
\orcid{0000-0002-6359-2774}
\affiliation{%
  \institution{Ubiquitous Knowledge Processing Lab (UKP)\\
  Technical University of Darmstadt}
  \city{Darmstadt}
  \country{Germany}
  }
\email{ilia.kuznetsov@tu-darmstadt.de}

\author{Florian Müller}
\orcid{0000-0002-9621-6214}
\affiliation{%
  \institution{Human-Computer Interaction\\
  Technical University of Darmstadt}
  \city{Darmstadt}
  \country{Germany}
  }
\email{florian.mueller@tu-darmstadt.de}

\author{Iryna Gurevych}
\orcid{0000-0003-2187-7621}
\affiliation{%
  \institution{Ubiquitous Knowledge Processing Lab (UKP)\\
  Technical University of Darmstadt}
  \city{Darmstadt}
  \country{Germany}
  }
\email{iryna.gurevych@tu-darmstadt.de}

\renewcommand{\shortauthors}{Zyska et. al.}

\begin{abstract}
Educational technologies often misalign with instructors' pedagogical goals, forcing adaptations that compromise teaching efficacy.
In this paper, we present a case study on the co-development of curriculum and technology in the context of a university course on scientific writing. Specifically, we examine how a custom-built peer feedback system was iteratively developed alongside the course to support annotation, feedback exchange, and revision.
Results show that while co-development fostered stronger alignment between software features and course goals, it also exposed usability limitations and infrastructure-related frustrations, emphasizing the need for closer coordination between teaching and technical teams.
\end{abstract}



\maketitle

\section{Introduction}
From assignment submissions to grading and communication, Learning Management Systems (LMS) such as Moodle or ILIAS have become a foundational infrastructure in higher education. These platforms are typically generic in design and functionality, offering a one-size-fits-all approach that often misaligns with the specific pedagogical goals of individual courses and educators~\cite{gamageSystematicReviewTrends2022, cabero-almenaraTechnicalDidacticKnowledge2019}. As a result, educators feel forced to adapt their teaching to fit predefined workflows, or to repurpose existing features to approximate what they actually want to do~\cite{chughImplementingEducationalTechnology2023}. Consequently, teaching and technology often feel misaligned for students and educators~\cite{luoSystematicReviewEvaluation2024}, creating inefficiencies, increasing administrative overhead, and limiting the positive impact of educational technology~\cite{akintayoEvaluatingImpactEducational2024}.

To overcome these challenges, prior work called for closer integration of pedagogy and technology through tandem design of software and curriculum~\cite{akbarDigitalTechnologyShaping2016, eadyToolsLearningTechnology2013a}. The promise of this approach lies in the idea that as pedagogical goals and software infrastructure evolve together, they can support coherent workflows, reduce friction, and preserve educators’ control over content delivery. 
However, we still lack insight into how such co-development plays out in practice, including the challenges and experiences of those involved.

In this paper, we explore what this process looks like in practice. We present a case study of a large university course on scientific writing integrating a custom peer feedback system. This system enables students and educators to annotate and review each other’s submissions using structured tags and free-text comments. Each student reviewed two peers and received both peer and educator feedback, which they used to revise their submission. The platform was developed iteratively during the course, adapting to legal, pedagogical, and infrastructural needs as they emerged. 

The contribution of this paper is two-fold: First, we show how pedagogical, technical, and organizational factors interact in the co-development of curriculum and educational software. Second, based on mixed-methods data, we analyze how course and platform co-evolved, revealing both the potential for better pedagogical alignment and tensions around feedback, usability, and AI integration.

\section{Background and Related Work}

\paragraph*{\textbf{Technology in Higher education}}
Implementation research \cite{Century2016} studies the extent to which software implementation efforts achieve the specified goals. Implementation research has a long history in EdTech; for example, in their analysis of 46 empirical research studies with focus on technology implementation issues, Chugh et al. \cite{Chugh2023} highlight significant challenges such as technology and stakeholder barriers \cite{Ranbir2024} or the ability of students and educators to cope with new technology \cite{Chugh2023}. While previous studies explored stakeholder experiences with \textit{existing} implementations or \textit{established} courses \cite{munday2016duolingo, Ryan2012, Jin2022}, our study focuses on how software and curriculum are co-developed to meet pedagogical goals.

\paragraph*{\textbf{Peer Feedback and Learning}}
Peer feedback is essential in higher education, as it helps students critique peer work \cite{topping1998peer} and reflect on their own writing \cite{cho2010student}, and supports learning at scale when providing direct instructor feedback is not feasible \cite{Søndergaard2012}. Peer feedback is widely used both in university courses and in online education, and is an active research area in educational psychology \cite{Bauer2023}.
To support peer feedback, digital tools like Juxtapeer \cite{Cambre2018} and PeerStudio \cite{Kulkarni2015PeerStudio} guide students through structured, rubric-based reviews. These systems are optimized for short-term, synchronous settings and support novel features such as comparing submissions side-by-side or routing them to active peers for quick turnaround. In contrast to these works, we explore a scenario of a semester-long university course with asynchronous participation, slower feedback cycles, and deeper curricular integration.

\paragraph*{\textbf{CARE}}
The \textbf{C}ollaborative \textbf{A}I-Assisted \textbf{R}eading \textbf{E}nvironment \cite{Zyska2023} was selected as the core platform for this course due to its unique capacity to support systematic data collection and workflow integration. Originally developed as a research tool, CARE’s architecture enables fine-grained text annotation with pre-defined semantic and color-coded tagging (i.e., Highlight, Strength, Weakness, Other) derived from Kuznetsov et. al. \cite{Kuznetsov2022}, and interaction logging. To operationalize CARE within a live university course, the platform had to be significantly extended, as the original scope did not fully address the pedagogical, procedural, and legal requirements of curricular integration.

\section{Course and software development}

\paragraph*{\textbf{Goals, curriculum and stakeholders.}}
Our study focuses on the Bachelor-level course “Introduction to Scientific Work” at the computer science department of a major European university, launched in Winter 2024. As part of the course,  \textsc{Students} write a six-page exposé on a predefined computer science topic. Each student provides two anonymous peer reviews, receives feedback from peers and \textsc{Educators}, and revises the exposé accordingly. \textsc{Developers} are responsible for the development and deployment of the software. The \textsc{Educators} included one lecturer, two doctoral researchers, one postdoctoral researcher, and six teaching assistants. \textsc{Developers} were a team of five research assistants led by a doctoral researcher. In total, 193 bachelor \textsc{Students} enrolled in the course; 159 completed the course.

\begin{figure*}[t]
  \centering
  \includegraphics[width=\textwidth]{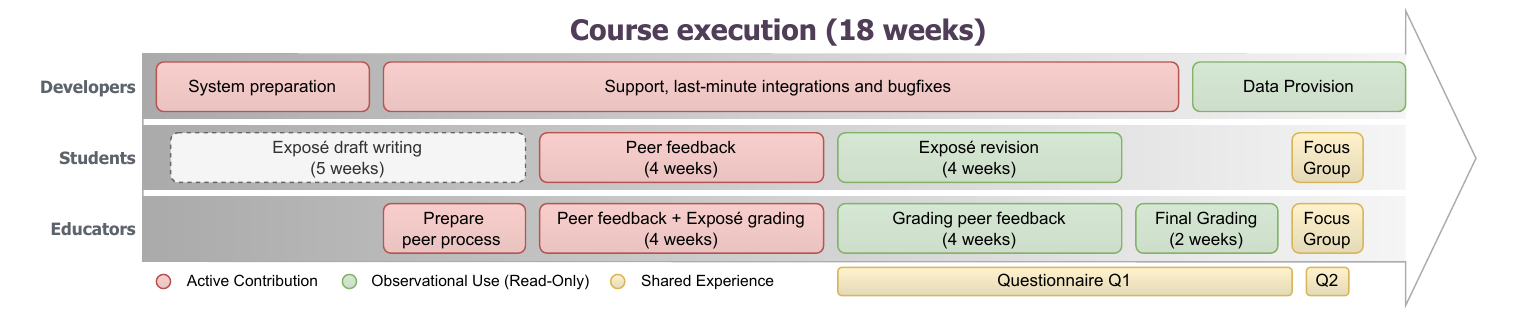}
  \caption{Touch-points of developers, educators, and students with CARE along the course. Color-coded regions indicate the stakeholder engagement with the system: \textit{red} areas denote active contributions, \textit{green} areas represent read-only or observational use, \textit{yellow} marks phases where stakeholders share their experience or feedback.}
  \Description{Timeline of developer, student, and educator engagement during an 18-week course, structured by stakeholder roles (developers, students, and educators), and color-coded by type of engagement: red for active contributions, green for observational use (read-only), and yellow for shared experience or feedback collection. Developers are actively involved in system preparation at the start, followed by continued support, last-minute integrations, and bug fixes during the course. They engage in observational data provision toward the end. Students begin with a five-week exposé draft writing phase, with no system interaction. This is followed by a four-week peer feedback period (active engagement), then a four-week exposé revision phase (read-only), and finally, a shared experience during the focus group sessions. Educators start by preparing the peer review process (active contribution), then manage peer feedback and exposé grading over four weeks. They also observe from the system the peer feedback grading for four weeks, and actively participate in final grading during the last two weeks. Educators also participate in shared experience activities in a subsequent focus group. The first questionnaires is indicated in the middle of the course and the second at the end of the course.}
  \label{fig:workflow_tasks}
\end{figure*}

\noindent\textbf{Stakeholder touch points and integration.} Each stakeholder group played a distinct role in integrating CARE into the course workflow. \textsc{Students} used CARE in two stages: first, to annotate and review the peers' exposés, summarizing their assessments in free-text form using previously introduced feedback principles and exposé-writing guidelines; then, to access feedback on their own work and revise accordingly. \textsc{Educators} handled both technical and pedagogical tasks: transferring exposés from Moodle, managing user credentials, assigning peers, and assessing and grading submissions using annotations and written feedback. Together with the \textsc{Developers}, they trained users and later extracted data for analysis. The \textsc{Developers} oversaw the development and deployment of CARE, integrated it with Moodle, ensured stable operation of the tool, provided support, and conducted the studies. 

\noindent\textbf{Mutual adaptation of course and software.} The course design directly influenced key software features. Embedding informed consent into the UI stemmed from the need to ensure ethical clarity at the point of student interaction. The data collection process through CARE required an ethics approval process, which in turn provided a concrete and timely case study for teaching research ethics, anchoring ethical education in the \textsc{Students'} direct experience. The demand for open-ended feedback pushed the implementation of a free-form editor, while the requirement for structured annotations and PDF-based submissions led to a mandatory LaTeX template, aligning student output with software processing needs. Contractually agreed working hour variations of the \textsc{Educators} shaped flexible reviewer assignment strategies. Uncertainties in availability and workload led to fallback mechanisms for reassigning review duties and prompted a role management logic. Pedagogical structure also dictated timing: lectures on feedback and writing were placed before introducing CARE, and synchronized with credential distribution. Institutional workflows demanded an integration with the Moodle API, enabling seamless data access and minimizing friction for \textsc{Students} and \textsc{Educators}. 

\section{Methodology}

To understand stakeholder experience with CARE and gather improvement suggestions, we distributed a questionnaire (\texttt{Q1}) to \textsc{Students}, followed by two focus groups \cite{Blandford2016}: one with teaching assistants from \textsc{Educators} group (Focus Group Educators, \texttt{FE}) and one with \textsc{students} (Focus Group Students, \texttt{FS}).

Questionnaire \texttt{Q1} was distributed shortly after the peer feedback phase via Moodle. The survey assessed usability using the Usability Metric for User Experience (UMUX) scale \cite{Finstad2010}.
It also included open-ended questions about user experiences and potential friction points when interacting with CARE. 

Focus groups \texttt{FE} and \texttt{FS} were conducted after the end of the course. Each session followed a semi-structured format.  
After giving informed consent, participants completed a warm-up questionnaire (\texttt{Q2}) covering demographics and attitudes toward digital learning technologies.

After this, participants discussed their experiences with CARE, guided by open-ended questions and followed by a brainwriting exercise \cite{Schlicksupp1976} to surface improvement ideas. Finally, participants prioritized these ideas using dot voting. \texttt{FS} took place on-site, while \texttt{FE} was conducted remotely using Zoom and a Miro Board. A researcher unaffiliated with the course moderated all focus group sessions. 

We recruited participants via direct outreach and Moodle; participation was voluntary. Two teaching assistants (1 man, 1 woman) joined \texttt{FE} through their employment; three students (1 man, 1 woman, 1 undisclosed) joined \texttt{FS} and were compensated at minimum wage. We transcribed recordings with Whisper \cite{radford2022}, anonymized and coded them in QualCoder.\footnote{\url{https://qualcoder.wordpress.com}} Three researchers independently coded and iteratively refined themes.

\section{Results}

We focus on usability, workflow integration, and satisfaction, and identify areas for improvement. We contextualize findings using participants' EdTech experience and tool expectations. As focus groups were held in German, we provide English translations.

\begin{figure}[tb]
  \centering
  \begin{subfigure}[t]{0.48\textwidth}
    \centering
    \includegraphics[width=\linewidth]{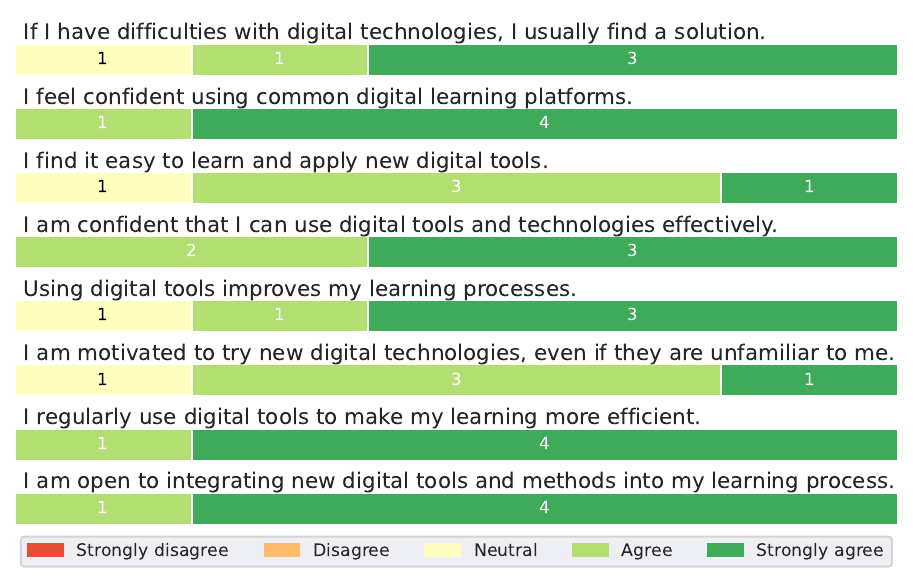}
    \caption{Digital Tool Acceptance (n=5)}
    \label{fig:q3_tool_acceptance}
  \end{subfigure}
  \hfill
  \begin{subfigure}[t]{0.48\textwidth}
    \centering
    \includegraphics[width=\linewidth]{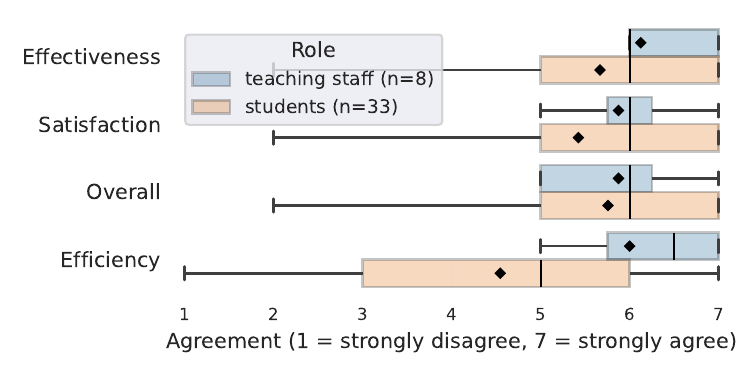}
    \caption{Usability Metric for User Experience (UMUX)}
    \label{fig:q1_umux}
  \end{subfigure}
  \caption{Summary of user responses: (a) Tool acceptance and (b) UMUX score.}
  \Description{Summary of user-reported digital tool acceptance and system usability experience, presented as two adjacent subfigures. The two related visualizations are shown side-by-side. (a) Digital Tool Acceptance (n=5): A horizontal bar chart displays the distribution of individual responses from five users to eight statements about digital tool usage in learning. Most responses are positive, indicating high confidence, motivation, and perceived benefits of using digital technologies. Color coding from red to green reflects disagreement to strong agreement. (b) Usability Metric for User Experience (UMUX): A grouped box plot compares ratings from teaching staff (n=8) and students (n=33) across four usability dimensions: effectiveness, satisfaction, overall experience, and efficiency. Teaching staff reported higher average scores, especially in efficiency and effectiveness. Students displayed more variation, particularly in efficiency.}
  \label{fig:side_by_side}
\end{figure}

\subsection{User Experience and System Perception}

No focus group participants had previously used dedicated peer feedback systems like CARE. Instead, they previously relied on general-purpose tools such as Adobe Acrobat Reader, Google Docs, or Overleaf to handle feedback tasks. Despite this unfamiliarity, overall responses were positive. In \texttt{Q1}, the UMUX (Figure~\ref{fig:q1_umux}) showed high satisfaction (M = 75.2, SD = 17.76, N = 39); \textsc{Educators} rated the system higher (M = 82.81, SD = 15.18, N = 8) than \textsc{students} (M = 73.19, SD = 18.06, N = 31), indicating a difference in perceived usability between user groups. In focus groups, \texttt{FE} participants expressed a strong interest in contributing to the tool's improvement, while \texttt{FS} participants were primarily motivated by a desire to enhance the overall course experience. Open-ended responses revealed that CARE was broadly perceived as fast, intuitive, and well-structured. Users appreciated its clear visual design and structured annotation workflow (Q1), particularly the pedagogically motivated use of pre-defined color-coded feedback tags that ``\textit{helped to think more systematically about what I was reviewing}'' (FS1). As reflected in Figure~\ref{fig:q3_tool_acceptance}, responses to tool acceptance items were predominantly positive, indicating a general willingness to engage with digital systems in the learning process. Yet, participants also reported challenges that affected efficiency and satisfaction, especially regarding CARE’s structuring of the feedback process and workflow.

\subsection{Effects on Workflow Efficiency}

CARE’s co-development with the course’s peer feedback component aimed to align pedagogical goals with technological functionality. Yet, real-world usage revealed frictions where course design and software logic misaligned. Initially, CARE implemented a linear review workflow that guided users step-by-step through the process. This structure conflicted with the course design that required flexible transitions between stages, prompting an interface update to support bidirectional navigation between the annotation view and free-form editor.
Though better aligned with course goals, it created practical challenges.
Users ``\textit{ended up opening two browser tabs [...] because there was no way to use annotations while writing the review}'' (\texttt{FS2}) and suggests that ``\textit{it would have helped to see annotations and write feedback side-by-side}'' (\texttt{FE1}). In particular, \textsc{Students} struggled with the navigation: they switched between views significantly more often (M = 13.49, SD = 11.69, n = 107) than \textsc{Educators} (M = 7.51, SD = 6.29, n = 173), t(144.5) = 4.86, p < .001 (Welch’s $t$-test). This behavior suggests that the system’s navigation still does not fully accommodate users’ natural feedback workflows.

Workflow was further impacted by CARE’s server-dependent architecture. While the system performed well under stable internet conditions, intermittent connectivity (e.g., during commute) led to synchronization failures, where participants ``\textit{lost parts of the review [...] when the connection dropped}'' (\texttt{FS2}).
``\textit{On the train}'', where ``\textit{the connection was bad}'', inputs were still accepted, but the lack of immediate error signaling led to user frustration, as ``\textit{it was annoying not knowing if anything worked}'' (\texttt{FE1}).
These limitations led to the emergence of parallel user workflows, such as writing ``\textit{the feedback in a separate file to avoid losing data}'' (\texttt{FS2}). The annotation system of CARE was praised for its tagging functionality and efficiency gains due to the reduction of manual input.
However, practical issues such as missing filtering functionality, limited editing and imprecise text selection caused confusion and impaired workflow efficiency. 

\subsection{Reflections on AI Integration}
Throughout both focus groups, one recurring topic was the potential role of AI within CARE. While not a core focus of our implementation, the use of AI in feedback sparked an active discussion. \textsc{Educators} favored AI assistance, including automated quality control, consistency checks, feedback generation and grading assistance. \textsc{Students} expressed more caution. While open to the use of AI for superficial tasks like grammar and style correction, or as a help with initial annotations, \textsc{Student} participants emphasized that ``\textit{the whole points is [...] to learn how to write feedback}'' (\texttt{FS3}) as a central learning experience which should not be fully automated as this would undermine a key pedagogical goal. The idea of AI-supported grading was particularly controversial. While both groups saw value in using AI for provisional suggestions, they emphasized that final grading decisions should remain with human instructors and ``\textit{the final call should always come from a real person}'' (\texttt{FE1}). Privacy also emerged as a key issue, with a clear preference for locally running models, as they ``\textit{don't want the data going who-knows-where}'' (\texttt{FS2}). Notably, despite the interest in AI, \textsc{Students} consistently expressed a different priority: ``\textit{Before adding smart features, just make the system more stable.}'' (\texttt{FS3}), meaning, resolving workflow frictions was more urgent than introducing new, intelligent tools.

\balance
\section{Discussion and Key Takeaways}

\paragraph*{\textbf{Software and curriculum co-evolve under constraints.}}
CARE had to align with institutional standards and the course’s pedagogical flow. Legal, technical, and scheduling constraints shaped features like access control and Moodle integration. 
In turn, these features influenced lecture sequencing and grading. The findings show that EdTech integration is not merely about fitting a tool into an existing curriculum but about co-developing technical and instructional frameworks under real-world constraints.
Successful implementations depend on tightly coupling software architecture with educational priorities, while remaining flexible to evolving institutional demands.

\paragraph*{\textbf{Software design should align to pedagogical practice.}}
\textsc{Students} and \textsc{Educators} valued CARE’s annotation features, especially the semantic tags that supported structured, theory-driven feedback with minimal manual effort. These design elements aligned well with pedagogical goals and contributed to a more focused review process.
In contrast, the system’s initial step-by-step workflow posed challenges. Although intended to guide users through a structured review sequence, it conflicted with how \textsc{students} naturally approached the task.
This points to a broader tension between formal instructional structures and the informal, often fluid nature of actual student behavior~\cite{Kinshuk2009}. While the design reflected pedagogical reasoning, it underestimated the cognitive overhead and disruption caused by rigid sequencing.
The results highlight that successful software integration depends not only on pedagogical alignment but also on sensitivity to user workflows.

\paragraph*{\textbf{Software reliability is central to workflow efficiency.}}
Users consistently described CARE as fast and responsive during regular use, which contributed to smooth task execution and a generally positive user experience. The system’s performance under stable conditions supported focused engagement, particularly during the annotation phase.
However, when connectivity issues occurred, even minor disruptions quickly impacted user engagement. \textsc{Students} and \textsc{Educators} often resorted to ad hoc parallel workflows using external tools, which complicated later reintegration and broke the intended workflow.
These workarounds point to a broader need for technical resilience in educational systems. 
Ensuring reliability through features like autosave, local caching, or offline access is essential to maintain workflow continuity, especially in real-world educational environments with variable access conditions.
 
\paragraph*{\textbf{A multiple-perspective approach is needed.}}
CARE's integration into peer review workflows revealed contrasting stakeholder expectations regarding feedback processes and automation.
\textsc{Educators} prioritized efficiency and standardization, welcoming AI-driven support to streamline grading and feedback generation.
\textsc{Students}, in contrast, valued peer review as a learning opportunity. They saw potential in AI as an assistive tool but resisted full automation, concerned about authenticity and skill development.
Close coordination between \textsc{Educators} and \textsc{Developers} is essential: \textsc{Developers} ensure system functionality, while \textsc{Educators} identify emerging issues early, through their direct contact with \textsc{students} and translate them into practical adjustments.
These divergent perspectives highlight the importance of designing EdTech systems that accommodate multiple user expectations and roles.

\section{Conclusion}

Rather than treating technology as a fixed solution, our case illustrates how curriculum and software can mutually shape each other. This co-development requires close collaboration between \textsc{developers} and \textsc{educators}, rapid iteration, and ongoing responsiveness to user experience. Our study has natural \textbf{limitations}, as our findings stem from a single University course with a limited number of focus group participants, with a focus on a particular peer feedback tool. Thus, one productive \textbf{future research} direction is to apply our proposed methodology to other courses, tools and environments, with particular attention given to refining navigation, improving feedback workflows, and minimizing friction during critical teaching phases. Another promising research direction is to explore the long-term impact of co-developed tools across diverse course formats and disciplines. Finally, there is room to examine how AI can assist without undermining key learning objectives, especially in feedback-intensive scenarios.

\paragraph*{\textbf{Implications.}} 
At a broader level, our findings call for a shift in how educational technology is conceived and deployed. Instead of adopting generic tools and adapting pedagogy around them, educators should embrace co-development practices that treat software as an evolving component of curriculum design. This requires institutional structures that support iterative development, cross-functional teams, and space for experimentation. Beyond higher education, this approach holds promise for any learning environment where feedback, transparency, and workflow alignment matter.




\begin{acks}
This work has been funded by the German Research Foundation (DFG) as part of the PEER project (grant GU 798/28-1) and by the European Union (ERC, InterText, 101054961). Views and opinions expressed are however those of the author(s) only and do not necessarily reflect those of the European Union or the European Research Council. Neither the European Union nor the granting authority can be held responsible for them.
\end{acks}

\bibliographystyle{ACM-Reference-Format}
\bibliography{sources}


\begin{thebibliography}{26}


\ifx \showCODEN    \undefined \def \showCODEN     #1{\unskip}     \fi
\ifx \showISBNx    \undefined \def \showISBNx     #1{\unskip}     \fi
\ifx \showISBNxiii \undefined \def \showISBNxiii  #1{\unskip}     \fi
\ifx \showISSN     \undefined \def \showISSN      #1{\unskip}     \fi
\ifx \showLCCN     \undefined \def \showLCCN      #1{\unskip}     \fi
\ifx \shownote     \undefined \def \shownote      #1{#1}          \fi
\ifx \showarticletitle \undefined \def \showarticletitle #1{#1}   \fi
\ifx \showURL      \undefined \def \showURL       {\relax}        \fi
\providecommand\bibfield[2]{#2}
\providecommand\bibinfo[2]{#2}
\providecommand\natexlab[1]{#1}
\providecommand\showeprint[2][]{arXiv:#2}

\bibitem[Akbar(2016)]%
        {akbarDigitalTechnologyShaping2016}
\bibfield{author}{\bibinfo{person}{Monika Akbar}.} \bibinfo{year}{2016}\natexlab{}.
\newblock \showarticletitle{Digital {{Technology Shaping Teaching Practices}} in {{Higher Education}}}.
\newblock \bibinfo{journal}{\emph{Frontiers in ICT}}  \bibinfo{volume}{3} (\bibinfo{year}{2016}).
\newblock
\showISSN{2297-198X}
\href{https://doi.org/10.3389/fict.2016.00001}{doi:\nolinkurl{10.3389/fict.2016.00001}}


\bibitem[Akintayo et~al\mbox{.}(2024)]%
        {akintayoEvaluatingImpactEducational2024}
\bibfield{author}{\bibinfo{person}{Olateju~Temitope Akintayo}, \bibinfo{person}{Chima~Abimbola Eden}, \bibinfo{person}{Oyebola~Olusola Ayeni}, {and} \bibinfo{person}{Nneamaka~Chisom Onyebuchi}.} \bibinfo{year}{2024}\natexlab{}.
\newblock \showarticletitle{Evaluating the Impact of Educational Technology on Learning Outcomes in the Higher Education Sector: A Systematic Review}.
\newblock \bibinfo{journal}{\emph{International Journal of Management \& Entrepreneurship Research}} \bibinfo{volume}{6}, \bibinfo{number}{5} (\bibinfo{year}{2024}), \bibinfo{pages}{1395--1422}.
\newblock
\showISSN{2664-3596}
\href{https://doi.org/10.51594/ijmer.v6i5.1091}{doi:\nolinkurl{10.51594/ijmer.v6i5.1091}}


\bibitem[Bauer et~al\mbox{.}(2023)]%
        {Bauer2023}
\bibfield{author}{\bibinfo{person}{Elisabeth Bauer}, \bibinfo{person}{Martin Greisel}, \bibinfo{person}{Ilia Kuznetsov}, \bibinfo{person}{Markus Berndt}, \bibinfo{person}{Ingo Kollar}, \bibinfo{person}{Markus Dresel}, \bibinfo{person}{Martin~R. Fischer}, {and} \bibinfo{person}{Frank Fischer}.} \bibinfo{year}{2023}\natexlab{}.
\newblock \showarticletitle{Using natural language processing to support peer-feedback in the age of artificial intelligence: A cross-disciplinary framework and a research agenda}.
\newblock \bibinfo{journal}{\emph{British Journal of Educational Technology}} \bibinfo{volume}{54}, \bibinfo{number}{5} (\bibinfo{year}{2023}), \bibinfo{pages}{1222--1245}.
\newblock
\href{https://doi.org/10.1111/bjet.13336}{doi:\nolinkurl{10.1111/bjet.13336}}
\showeprint{https://bera-journals.onlinelibrary.wiley.com/doi/pdf/10.1111/bjet.13336}


\bibitem[Blandford et~al\mbox{.}(2016)]%
        {Blandford2016}
\bibfield{author}{\bibinfo{person}{Ann Blandford}, \bibinfo{person}{Dominic Furniss}, {and} \bibinfo{person}{Stephann Makri}.} \bibinfo{year}{2016}\natexlab{}.
\newblock \bibinfo{booktitle}{\emph{Qualitative {HCI} Research: Going Behind the Scenes}}.
\newblock \bibinfo{publisher}{Morgan {\&} Claypool Publishers}, \bibinfo{address}{San Rafael, California]}.
\newblock
\showISBNx{978-3-031-01089-7}
\href{https://doi.org/10.2200/S00706ED1V01Y201602HCI034}{doi:\nolinkurl{10.2200/S00706ED1V01Y201602HCI034}}


\bibitem[{Cabero-Almenara} et~al\mbox{.}(2019)]%
        {cabero-almenaraTechnicalDidacticKnowledge2019}
\bibfield{author}{\bibinfo{person}{Julio {Cabero-Almenara}}, \bibinfo{person}{Maria~Luisa Arancibia}, {and} \bibinfo{person}{Annachiara Del~Prete}.} \bibinfo{year}{2019}\natexlab{}.
\newblock \showarticletitle{Technical and {{Didactic Knowledge}} of the {{Moodle LMS}} in {{Higher Education}}. {{Beyond Functional Use}}}.
\newblock \bibinfo{journal}{\emph{Journal of New Approaches in Educational Research}} \bibinfo{volume}{8}, \bibinfo{number}{1} (\bibinfo{year}{2019}), \bibinfo{pages}{25--33}.
\newblock
\showISSN{2254-7339}
\href{https://doi.org/10.7821/naer.2019.1.327}{doi:\nolinkurl{10.7821/naer.2019.1.327}}


\bibitem[Cambre et~al\mbox{.}(2018)]%
        {Cambre2018}
\bibfield{author}{\bibinfo{person}{Julia Cambre}, \bibinfo{person}{Scott Klemmer}, {and} \bibinfo{person}{Chinmay Kulkarni}.} \bibinfo{year}{2018}\natexlab{}.
\newblock \showarticletitle{Juxtapeer: Comparative Peer Review Yields Higher Quality Feedback and Promotes Deeper Reflection}. In \bibinfo{booktitle}{\emph{Proceedings of the 2018 CHI Conference on Human Factors in Computing Systems}} (Montreal QC, Canada) \emph{(\bibinfo{series}{CHI '18})}. \bibinfo{publisher}{Association for Computing Machinery}, \bibinfo{address}{New York, NY, USA}, \bibinfo{pages}{1–13}.
\newblock
\showISBNx{9781450356206}
\href{https://doi.org/10.1145/3173574.3173868}{doi:\nolinkurl{10.1145/3173574.3173868}}


\bibitem[Century and Cassata(2016)]%
        {Century2016}
\bibfield{author}{\bibinfo{person}{Jeanne Century} {and} \bibinfo{person}{Amy Cassata}.} \bibinfo{year}{2016}\natexlab{}.
\newblock \showarticletitle{Implementation Research: Finding Common Ground on What, How, Why, Where, and Who}.
\newblock \bibinfo{journal}{\emph{Review of Research in Education}} \bibinfo{volume}{40}, \bibinfo{number}{1} (\bibinfo{year}{2016}), \bibinfo{pages}{169--215}.
\newblock
\href{https://doi.org/10.3102/0091732X16665332}{doi:\nolinkurl{10.3102/0091732X16665332}}
\showeprint{https://doi.org/10.3102/0091732X16665332}


\bibitem[Cho and MacArthur(2010)]%
        {cho2010student}
\bibfield{author}{\bibinfo{person}{Kwangsu Cho} {and} \bibinfo{person}{Charles MacArthur}.} \bibinfo{year}{2010}\natexlab{}.
\newblock \showarticletitle{Student revision with peer and expert reviewing}.
\newblock \bibinfo{journal}{\emph{Learning and instruction}} \bibinfo{volume}{20}, \bibinfo{number}{4} (\bibinfo{year}{2010}), \bibinfo{pages}{328--338}.
\newblock


\bibitem[Chugh et~al\mbox{.}(2023a)]%
        {chughImplementingEducationalTechnology2023}
\bibfield{author}{\bibinfo{person}{Ritesh Chugh}, \bibinfo{person}{Darren Turnbull}, \bibinfo{person}{Michael~A. Cowling}, \bibinfo{person}{Robert Vanderburg}, {and} \bibinfo{person}{Michelle~A. Vanderburg}.} \bibinfo{year}{2023}\natexlab{a}.
\newblock \showarticletitle{Implementing Educational Technology in {{Higher Education Institutions}}: {{A}} Review of Technologies, Stakeholder Perceptions, Frameworks and Metrics}.
\newblock \bibinfo{journal}{\emph{Education and Information Technologies}} \bibinfo{volume}{28}, \bibinfo{number}{12} (\bibinfo{year}{2023}), \bibinfo{pages}{16403--16429}.
\newblock
\showISSN{1573-7608}
\href{https://doi.org/10.1007/s10639-023-11846-x}{doi:\nolinkurl{10.1007/s10639-023-11846-x}}


\bibitem[Chugh et~al\mbox{.}(2023b)]%
        {Chugh2023}
\bibfield{author}{\bibinfo{person}{Ritesh Chugh}, \bibinfo{person}{Darren Turnbull}, \bibinfo{person}{Michael~A. Cowling}, \bibinfo{person}{Robert Vanderburg}, {and} \bibinfo{person}{Michelle~A. Vanderburg}.} \bibinfo{year}{2023}\natexlab{b}.
\newblock \showarticletitle{Implementing educational technology in Higher Education Institutions: A review of technologies, stakeholder perceptions, frameworks and metrics}.
\newblock \bibinfo{journal}{\emph{Education and Information Technologies}} \bibinfo{volume}{28}, \bibinfo{number}{12} (\bibinfo{year}{2023}), \bibinfo{pages}{16403--16429}.
\newblock
\showISSN{1573-7608}
\href{https://doi.org/10.1007/s10639-023-11846-x}{doi:\nolinkurl{10.1007/s10639-023-11846-x}}


\bibitem[Eady and Lockyer(2013)]%
        {eadyToolsLearningTechnology2013a}
\bibfield{author}{\bibinfo{person}{Michelle Eady} {and} \bibinfo{person}{Lori Lockyer}.} \bibinfo{year}{2013}\natexlab{}.
\newblock \bibinfo{booktitle}{\emph{Tools for learning: technology and teaching strategies}}.
\newblock \bibinfo{publisher}{Cambridge University Press (CUP)}, \bibinfo{address}{United Kingdom}, \bibinfo{pages}{71--89}.
\newblock
\showISBNx{9781107672826}


\bibitem[Finstad(2010)]%
        {Finstad2010}
\bibfield{author}{\bibinfo{person}{Kraig Finstad}.} \bibinfo{year}{2010}\natexlab{}.
\newblock \showarticletitle{The Usability Metric for User Experience}.
\newblock \bibinfo{journal}{\emph{Interacting with Computers}}  \bibinfo{volume}{22} (\bibinfo{year}{2010}), \bibinfo{pages}{323--327}.
\newblock
\href{https://doi.org/10.1016/j.intcom.2010.04.004}{doi:\nolinkurl{10.1016/j.intcom.2010.04.004}}


\bibitem[Gamage et~al\mbox{.}(2022)]%
        {gamageSystematicReviewTrends2022}
\bibfield{author}{\bibinfo{person}{Sithara H. P.~W. Gamage}, \bibinfo{person}{Jennifer~R. Ayres}, {and} \bibinfo{person}{Monica~B. Behrend}.} \bibinfo{year}{2022}\natexlab{}.
\newblock \showarticletitle{A Systematic Review on Trends in Using {{Moodle}} for Teaching and Learning}.
\newblock \bibinfo{journal}{\emph{International Journal of STEM Education}} \bibinfo{volume}{9}, \bibinfo{number}{1} (\bibinfo{year}{2022}), \bibinfo{pages}{9}.
\newblock
\showISSN{2196-7822}
\href{https://doi.org/10.1186/s40594-021-00323-x}{doi:\nolinkurl{10.1186/s40594-021-00323-x}}


\bibitem[Jin et~al\mbox{.}(2022)]%
        {Jin2022}
\bibfield{author}{\bibinfo{person}{Qiao Jin}, \bibinfo{person}{Yu Liu}, \bibinfo{person}{Svetlana Yarosh}, \bibinfo{person}{Bo Han}, {and} \bibinfo{person}{Feng Qian}.} \bibinfo{year}{2022}\natexlab{}.
\newblock \showarticletitle{How Will VR Enter University Classrooms? Multi-stakeholders Investigation of VR in Higher Education}. In \bibinfo{booktitle}{\emph{Proceedings of the 2022 CHI Conference on Human Factors in Computing Systems}} (New Orleans, LA, USA) \emph{(\bibinfo{series}{CHI '22})}. \bibinfo{publisher}{Association for Computing Machinery}, \bibinfo{address}{New York, NY, USA}, Article \bibinfo{articleno}{563}, \bibinfo{numpages}{17}~pages.
\newblock
\showISBNx{9781450391573}
\href{https://doi.org/10.1145/3491102.3517542}{doi:\nolinkurl{10.1145/3491102.3517542}}


\bibitem[{Kinshuk} et~al\mbox{.}(2009)]%
        {Kinshuk2009}
\bibfield{author}{\bibinfo{person}{{Kinshuk}}, \bibinfo{person}{Tzu-Chien Liu}, {and} \bibinfo{person}{Sabine Graf}.} \bibinfo{year}{2009}\natexlab{}.
\newblock \showarticletitle{Coping with mismatched courses: students' behaviour and performance in courses mismatched to their learning styles}.
\newblock \bibinfo{journal}{\emph{Educational Technology Research and Development}} \bibinfo{volume}{57}, \bibinfo{number}{6} (\bibinfo{year}{2009}), \bibinfo{pages}{739--752}.
\newblock
\showISSN{1556-6501}
\href{https://doi.org/10.1007/s11423-009-9116-y}{doi:\nolinkurl{10.1007/s11423-009-9116-y}}


\bibitem[Kulkarni et~al\mbox{.}(2015)]%
        {Kulkarni2015PeerStudio}
\bibfield{author}{\bibinfo{person}{Chinmay~E. Kulkarni}, \bibinfo{person}{Michael~S. Bernstein}, {and} \bibinfo{person}{Scott~R. Klemmer}.} \bibinfo{year}{2015}\natexlab{}.
\newblock \showarticletitle{PeerStudio: Rapid Peer Feedback Emphasizes Revision and Improves Performance}. In \bibinfo{booktitle}{\emph{Proceedings of the Second (2015) ACM Conference on Learning @ Scale}} (Vancouver, BC, Canada) \emph{(\bibinfo{series}{L@S '15})}. \bibinfo{publisher}{ACM}, \bibinfo{address}{New York, NY, USA}, \bibinfo{pages}{75--84}.
\newblock
\showISBNx{978-1-4503-3411-2}
\href{https://doi.org/10.1145/2724660.2724670}{doi:\nolinkurl{10.1145/2724660.2724670}}


\bibitem[Kuznetsov et~al\mbox{.}(2022)]%
        {Kuznetsov2022}
\bibfield{author}{\bibinfo{person}{Ilia Kuznetsov}, \bibinfo{person}{Jan Buchmann}, \bibinfo{person}{Max Eichler}, {and} \bibinfo{person}{Iryna Gurevych}.} \bibinfo{year}{2022}\natexlab{}.
\newblock \showarticletitle{Revise and Resubmit: An Intertextual Model of Text-based Collaboration in Peer Review}.
\newblock \bibinfo{journal}{\emph{Computational Linguistics}} \bibinfo{volume}{48}, \bibinfo{number}{4} (\bibinfo{year}{2022}), \bibinfo{pages}{949--986}.
\newblock
\href{https://doi.org/10.1162/coli_a_00455}{doi:\nolinkurl{10.1162/coli_a_00455}}


\bibitem[Luo et~al\mbox{.}(2024)]%
        {luoSystematicReviewEvaluation2024}
\bibfield{author}{\bibinfo{person}{Zhimin Luo}, \bibinfo{person}{Babar~Nawaz Abbasi}, \bibinfo{person}{Chong Yang}, \bibinfo{person}{Jiayin Li}, {and} \bibinfo{person}{Ali Sohail}.} \bibinfo{year}{2024}\natexlab{}.
\newblock \showarticletitle{A Systematic Review of Evaluation and Program Planning Strategies for Technology Integration in Education: {{Insights}} for Evidence-Based Practice}.
\newblock \bibinfo{journal}{\emph{Education and Information Technologies}} \bibinfo{volume}{29}, \bibinfo{number}{16} (\bibinfo{year}{2024}), \bibinfo{pages}{21133--21167}.
\newblock
\showISSN{1573-7608}
\href{https://doi.org/10.1007/s10639-024-12707-x}{doi:\nolinkurl{10.1007/s10639-024-12707-x}}


\bibitem[Munday(2016)]%
        {munday2016duolingo}
\bibfield{author}{\bibinfo{person}{Pilar Munday}.} \bibinfo{year}{2016}\natexlab{}.
\newblock \showarticletitle{The Case for Using DUOLINGO as Part of the Language Classroom Experience}.
\newblock \bibinfo{journal}{\emph{RIED. Revista Iberoamericana de Educación a Distancia}} \bibinfo{volume}{19}, \bibinfo{number}{1} (\bibinfo{year}{2016}), \bibinfo{pages}{83--101}.
\newblock
\href{https://doi.org/10.5944/ried.19.1.14581}{doi:\nolinkurl{10.5944/ried.19.1.14581}}


\bibitem[Radford et~al\mbox{.}(2023)]%
        {radford2022}
\bibfield{author}{\bibinfo{person}{Alec Radford}, \bibinfo{person}{Jong~Wook Kim}, \bibinfo{person}{Tao Xu}, \bibinfo{person}{Greg Brockman}, \bibinfo{person}{Christine McLeavey}, {and} \bibinfo{person}{Ilya Sutskever}.} \bibinfo{year}{2023}\natexlab{}.
\newblock \showarticletitle{Robust speech recognition via large-scale weak supervision}. In \bibinfo{booktitle}{\emph{Proceedings of the 40th International Conference on Machine Learning}} \emph{(\bibinfo{series}{ICML'23})}. \bibinfo{publisher}{JMLR.org}, \bibinfo{address}{Honolulu, Hawaii, USA}, Article \bibinfo{articleno}{1182}, \bibinfo{numpages}{27}~pages.
\newblock


\bibitem[Ranbir(2024)]%
        {Ranbir2024}
\bibfield{author}{\bibinfo{person}{Dr Ranbir}.} \bibinfo{year}{2024}\natexlab{}.
\newblock \showarticletitle{Educational Technology Integration: Challenges and Opportunities}.
\newblock \bibinfo{journal}{\emph{Innovative Research Thoughts}}  \bibinfo{volume}{10} (\bibinfo{year}{2024}), \bibinfo{pages}{75--79}.
\newblock
\href{https://doi.org/10.36676/irt.v10.i2.1406}{doi:\nolinkurl{10.36676/irt.v10.i2.1406}}


\bibitem[Schlicksupp(1976)]%
        {Schlicksupp1976}
\bibfield{author}{\bibinfo{person}{Helmut Schlicksupp}.} \bibinfo{year}{1976}\natexlab{}.
\newblock \bibinfo{booktitle}{\emph{Kreative Ideenfindung in der Unternehmung; Methoden und Modelle}}.
\newblock \bibinfo{publisher}{De Gruyter}, \bibinfo{address}{Berlin, Boston}.
\newblock
\showISBNx{9783110845044}
\href{https://doi.org/10.1515/9783110845044}{doi:\nolinkurl{10.1515/9783110845044}}


\bibitem[Søndergaard and and(2012)]%
        {Søndergaard2012}
\bibfield{author}{\bibinfo{person}{Harald Søndergaard} {and} \bibinfo{person}{Raoul A.~Mulder and}.} \bibinfo{year}{2012}\natexlab{}.
\newblock \showarticletitle{Collaborative learning through formative peer review: pedagogy, programs and potential}.
\newblock \bibinfo{journal}{\emph{Computer Science Education}} \bibinfo{volume}{22}, \bibinfo{number}{4} (\bibinfo{year}{2012}), \bibinfo{pages}{343--367}.
\newblock
\href{https://doi.org/10.1080/08993408.2012.728041}{doi:\nolinkurl{10.1080/08993408.2012.728041}}
\showeprint{https://doi.org/10.1080/08993408.2012.728041}


\bibitem[Thomas G~Ryan and Park(2012)]%
        {Ryan2012}
\bibfield{author}{\bibinfo{person}{Kyle~Charron Thomas G~Ryan, Mary~Toye} {and} \bibinfo{person}{Gavin Park}.} \bibinfo{year}{2012}\natexlab{}.
\newblock \showarticletitle{Learning Management System Migration: An Analysis of Stakeholder Perspectives}.
\newblock \bibinfo{journal}{\emph{International Review of Research in Open and Distributed Learning}} \bibinfo{volume}{13}, \bibinfo{number}{1} (\bibinfo{year}{2012}), \bibinfo{pages}{220--237}.
\newblock
\href{https://doi.org/10.19173/irrodl.v13i1.1126}{doi:\nolinkurl{10.19173/irrodl.v13i1.1126}}


\bibitem[Topping(1998)]%
        {topping1998peer}
\bibfield{author}{\bibinfo{person}{Keith Topping}.} \bibinfo{year}{1998}\natexlab{}.
\newblock \showarticletitle{Peer assessment between students in colleges and universities}.
\newblock \bibinfo{journal}{\emph{Review of educational Research}} \bibinfo{volume}{68}, \bibinfo{number}{3} (\bibinfo{year}{1998}), \bibinfo{pages}{249--276}.
\newblock


\bibitem[Zyska et~al\mbox{.}(2023)]%
        {Zyska2023}
\bibfield{author}{\bibinfo{person}{Dennis Zyska}, \bibinfo{person}{Nils Dycke}, \bibinfo{person}{Jan Buchmann}, \bibinfo{person}{Ilia Kuznetsov}, {and} \bibinfo{person}{Iryna Gurevych}.} \bibinfo{year}{2023}\natexlab{}.
\newblock \showarticletitle{{CARE}: Collaborative {AI}-Assisted Reading Environment}. In \bibinfo{booktitle}{\emph{Proceedings of the 61st Annual Meeting of the Association for Computational Linguistics (Volume 3: System Demonstrations)}}, \bibfield{editor}{\bibinfo{person}{Danushka Bollegala}, \bibinfo{person}{Ruihong Huang}, {and} \bibinfo{person}{Alan Ritter}} (Eds.). \bibinfo{publisher}{Association for Computational Linguistics}, \bibinfo{address}{Toronto, Canada}, \bibinfo{pages}{291--303}.
\newblock
\href{https://doi.org/10.18653/v1/2023.acl-demo.28}{doi:\nolinkurl{10.18653/v1/2023.acl-demo.28}}


\end{thebibliography}

\appendix

\end{document}